\documentclass[aps,prl,twocolumn,amsmath,floatfix,showpacs,superscriptaddress]{revtex4}

\usepackage{graphicx}
\usepackage{color}
\usepackage{marvosym}
\usepackage{wasysym}
\usepackage{pifont}
\usepackage{amssymb}
\usepackage{amsmath}
\usepackage{natbib}
\usepackage{amsfonts}
\usepackage{epsfig}
\usepackage[american]{babel}
\usepackage{bm} %boldmath

\newcommand{\br}{{\bf r}}

\newcommand{\etal}{{\emph{et al.} }}

\begin{document}

\title{Size and dynamics of vortex dipoles in dilute Bose--Einstein condensates}
\author{Pekko Kuopanportti}
\email{pekko.kuopanportti@tkk.fi}
\affiliation{Department of Applied Physics/COMP, Aalto University, P.O. Box 14100, FI-00076 AALTO, Finland}
\author{Jukka~A.~M.~Huhtam\"aki}
\affiliation{Department of Applied Physics/COMP, Aalto
University, P.O. Box 14100, FI-00076 AALTO, Finland}
\author{Mikko M\"ott\"onen}
\affiliation{Department of Applied Physics/COMP, Aalto University, P.O. Box 14100, FI-00076 AALTO, Finland}
\affiliation{Low Temperature Laboratory, Aalto University, P.O. Box 13500, FI-00076 AALTO, Finland}

\begin{abstract}
Recently, Freilich~\etal [Science {\bf 329}, 1182 (2010)] experimentally discovered stationary states of vortex dipoles, pairs of vortices of opposite circulation, in dilute Bose--Einstein condensates. To explain their observations, we perform simulations based on the Gross--Pitaevskii equation and obtain excellent quantitative agreement on the size of the stationary dipole. We also investigate how their imaging method, in which atoms are repeatedly extracted from a single condensate, affects the vortex dynamics. We find that it mainly induces isotropic size oscillations of the condensate without otherwise disturbing the vortex trajectories. Thus, the imaging technique appears to be a promising tool for studying real-time superfluid dynamics.
\end{abstract}



\maketitle

\emph{Introduction}---Quantized vortices are topological defects that carry angular momentum and often serve as conclusive evidence for superfluidity. A vortex dipole, on the other hand, consists of a pair of vortices of opposite circulation and can be considered as a basic topological structure that carries linear momentum. Although vortex dipoles are widespread in classical fluid flows~\cite{Voropayev1994}, their role in superfluids seems less well established. Given the appearance of vortices and antivortices in the Berezinskii--Kosterlitz--Thouless transition~\cite{BKT} and superfluid turbulence~\cite{turbulence}, a detailed study of vortex dipoles will contribute to a broader understanding of superfluid phenomena. 

Dilute Bose--Einstein condensates (BECs) are an ideal choice for investigating superfluid vortex dynamics~\cite{Fetter2009}, since they can be described by tractable theories~\cite{Dalfovo1999} and are highly controllable in experiments~\cite{Anderson2010}. Vortex dipoles have been addressed in various theoretical studies of dilute BECs~\cite{Martikainen2001,Schulte2002,Zhou2004,Crasovan2002,Crasovan2003,Mottonen2005,Pietila2006,Middelkamp2010,Stockhofe2010,Li2008,Klein2007,Geurts2008}, and stationary vortex dipoles~\cite{Crasovan2002,Crasovan2003,Mottonen2005,Pietila2006,Middelkamp2010,Stockhofe2010,Li2008} have been found to exist over a wide range of particle numbers and trap geometries. Other vortex cluster configurations that are stationary in nonrotated BECs, such as the vortex tripole and vortex quadrupole, have also been predicted~\cite{Crasovan2002,Crasovan2003,Mottonen2005,Pietila2006,Middelkamp2010,Stockhofe2010}. In recent experiments~\cite{Neely2010,Freilich2010}, the formation and dynamics of vortex dipoles were investigated in ${}^{87}$Rb condensates. Vortex tripoles have also been observed experimentally~\cite{Seman2010}.

Typically, vortices in dilute BECs are imaged by removing the trapping potential and expanding the condensate in order to make the vortex cores optically resolvable. However, the expansion prevents taking multiple images from a single BEC. In Ref.~\cite{Freilich2010}, Freilich \etal present a method of vortex probing that allows vortex dynamics to be observed in real time within a single BEC rather than by reconstruction of images from multiple BECs. The method is based on repeated extraction, expansion, and imaging of small fractions of the BEC. In the demonstration of the technique, vortices were spontaneously created during evaporative cooling due to the Kibble--Zurek mechanism, and the orbital dynamics of single vortices and vortex dipoles were observed with high precision. In particular, clear experimental evidence for stationary vortex dipoles was obtained.

In this Rapid Communication, we present numerical results based on the zero-temperature Gross--Pitaevskii equation to explain the experimental observations of Freilich \etal \cite{Freilich2010}.  Previously published predictions~\cite{Mottonen2005,Pietila2006,Middelkamp2010,Stockhofe2010,Li2008} for the properties of stationary vortex dipoles concern only very weakly interacting or two-dimensional systems and hence cannot be used for direct comparison with the experiment. Here, we present stationary vortex-dipole states for a BEC identical to the one used in Ref.~\cite{Freilich2010} and find excellent quantitative agreement with the experiment. In order to analyze the accuracy of the multishot imaging method, we also study the effect of the repeated removal of particles on the dynamics of the vortex dipoles, and observe that it excites mainly the breathing mode of the BEC but does not significantly disturb the vortex trajectories.

\emph{Theory and methods}---In the zero-temperature limit, the dynamics of a dilute BEC is accurately described by the Gross--Pitaevskii~(GP) equation~\cite{Dalfovo1999} for the condensate order parameter $\Psi$,
\begin{equation}\label{eq:GPE}
i\hbar\partial_t\Psi(\br,t) = \left[ {\cal H}+ g |\Psi(\br,t)|^2 \right] \Psi(\br,t),
\end{equation}
where the single-particle Hamiltonian ${\cal H}$ consists in our case of the kinetic energy operator and an axisymmetric harmonic trapping potential,
\begin{equation}\label{eq:hamiltonian}
{\cal H} = -\frac{\hbar^2}{2m}\nabla^2 + \frac{1}{2}m\omega_r^2\left( x^2+y^2\right) + \frac{1}{2}m\omega_z^2 z^2.
\end{equation}
The atom--atom interaction strength $g$ is related to the $s$-wave scattering length $a$ and the number of atoms $N$ by $g=4\pi\hbar^2aN/m$. Here, $m$ is the atomic mass and $\Psi$ is normalized such that $\int|\Psi|^2 d^3r=1$. Often, it is relevant to consider the dynamics only in the $xy$ plane and factor out the $z$ dependence of the order parameter. The general $z$-factorized form can be written as $\Psi(\br,t)=\psi(x,y,t)\zeta(z)\exp\left(-i \epsilon_z t/\hbar\right)$, where the function $\zeta$ is required to satisfy $\int|\zeta|^2 dz = 1$ and $\epsilon_z$ is a constant energy shift. By inserting this into Eq.~(\ref{eq:GPE}), multiplying by $\zeta^\ast \exp\left(i\epsilon_z t/\hbar\right)$, and integrating over $z$, we obtain the effectively two-dimensional (2D) GP equation
\begin{equation}\label{eq:GPE2}
i\hbar\partial_t\psi = \left[-\frac{\hbar^2\nabla_{2\mathrm{D}}^2}{2m} + \frac{1}{2}m\omega_r^2\left(x^2+y^2\right)+ g_{2\mathrm{D}} |\psi|^2 \right] \psi,
\end{equation}
where $g_{\mathrm{2D}}=g\int |\zeta|^4 dz$ is the reduced interaction strength and $\psi$ is normalized according to $\int |\psi|^2 d^2 r = 1$. The energy shift is set to $\epsilon_z = \int\left( m\omega_z^2 z^2 |\zeta|^2-\hbar^2 \zeta^\ast \zeta''/m\right)dz/2$. Although the $z$ dependence of the order parameter does not separate ideally in the configuration of Ref.~\cite{Freilich2010}, this technique still provides a useful model for vortex motion, as shown in the following.

Stationary states of the system satisfy the time-independent GP equation, which is obtained from Eq.~(\ref{eq:GPE}) or Eq.~(\ref{eq:GPE2}) with the replacement $i\hbar\partial_t \longrightarrow \mu$, where $\mu$ is the chemical potential. For sufficiently large $N$, a good approximation for the lowest-energy stationary state may be obtained by neglecting the kinetic-energy term in the GP equation. In Eq.~(\ref{eq:GPE2}), this Thomas--Fermi (TF) approximation yields the solution
\begin{equation}\label{eq:TF}
\psi_\mathrm{TF}(x,y)= \mathrm{Re}\left\{\sqrt{n_0\left[1-\left(x^2+y^2\right)/R_\mathrm{TF}^2\right]}\right\},
\end{equation}
where $n_0=\sqrt{m\omega_r^2/\pi g_{\mathrm{2D}}}$ is the peak density and the TF radius of the 2D BEC is given by $R_\mathrm{TF}=\sqrt[4]{4g_{\mathrm{2D}}/\pi m\omega_r^2}$. 

In Ref.~\cite{Freilich2010}, the dynamics of a BEC was observed by repeatedly extracting, expanding, and imaging small fractions of the trapped condensate across its entire spatial extent. The atoms were extracted by a brief (1--3~$\mathrm{\mu}$s) microwave pulse that transferred them to a magnetically untrapped state, creating a representation of the trapped cloud at that instant. We assume that the microwave pulse samples the BEC uniformly throughout its entire volume, extracting a fraction $p$ of the atoms at each spatial point, and neglect possible interactions between trapped and untrapped atoms. Under these circumstances, the effect of the pulse on the trapped condensate is to reduce the number of particles $N$ by the fraction $p$. In Eq.~(\ref{eq:GPE2}), this implies that the interaction parameter $g_{2\mathrm{D}}$ becomes time dependent,
\begin{equation}\label{eq:g2d}
g_{2\mathrm{D}}(t)=g_{2\mathrm{D}}(0)\left(1-p\right)^{\sum_j\Theta(t-t_j)},
\end{equation}
where $\Theta$ is the Heaviside step function and $\left\{t_j\right\}$ are the time instants of the microwave pulses. 

In the simulations, we search for stationary vortex-dipole states in three dimensions and study the dynamics of symmetric and asymmetric vortex dipoles in the effectively 2D limit when particles are gradually removed from the trap. 
The stationary vortex dipoles are solved from the three-dimensional time-independent GP equation by using a relaxation method, and the time-evolution is solved numerically from Eq.~(\ref{eq:GPE2}) by using a combination of operator splitting and the Crank--Nicolson scheme. To obtain dimensionless quantities, we measure length in units of the radial harmonic oscillator length $a_r=\sqrt{\hbar/m\omega_r}$, energy in units of $\hbar\omega_r$, and time in units of $1/\omega_r$. This results in dimensionless forms of Eqs.~(\ref{eq:GPE}) and (\ref{eq:GPE2}), where the interaction strengths are replaced with $\tilde{g}=g/\hbar\omega_r a_r^3=4 \pi N a / a_r$ and $\tilde{g}_\mathrm{2D}= g_{2\mathrm{D}}/\hbar\omega_r a_r^2$. We choose the simulation parameters according to Ref.~\cite{Freilich2010}, setting $\left(\omega_r,\omega_z\right)=2\pi\times\left(35.8,101.2\right)\,\mathrm{Hz}$ and using the values $a=5.31\textrm{ nm}$~\cite{Kempen2002} and $m=1.44\times 10^{-25}\textrm{ kg}$ corresponding to ${}^{87}\rm{Rb}$. Thus, we have $a_r=1.80\ \mu\mathrm{m}$ and $\tilde{g}= 0.0370 \times N$.

\begin{figure*}[tbh]
\includegraphics[
  width=345pt,
  keepaspectratio]{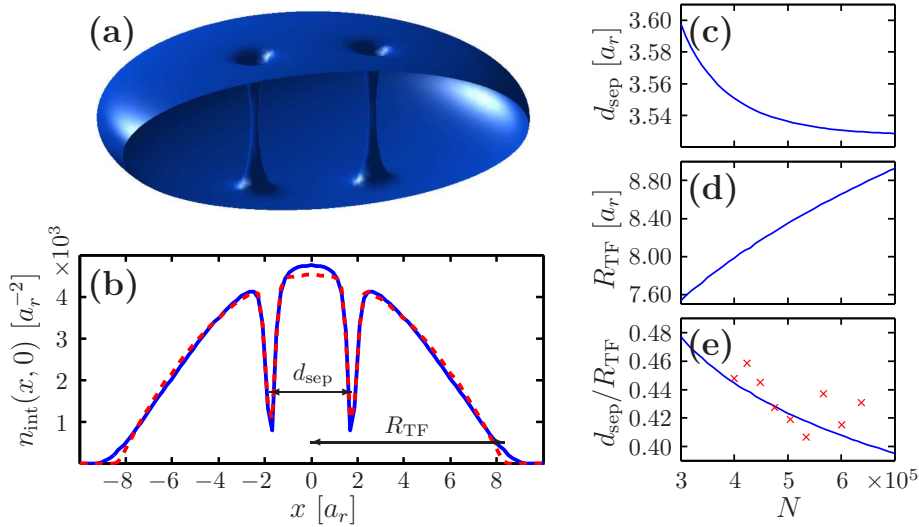}
\caption{\label{fig:statdip} (Color online) Stationary vortex dipole in three dimensions. (a) Isosurface of the particle density $|\Psi(\br)|^2$ for the stationary vortex dipole corresponding to the particle number $N=5 \times 10^5$. (b) Solid blue curve shows the $z$-integrated density along the $x$ axis, $n_{\mathrm{int}}(x,0)$, for the state in (a), whereas dashed red curve is a fit consisting of a sum of the TF profile [Eq.~(\ref{eq:TF})] and two Gaussians. Here, $d_\mathrm{sep}$ is the separation distance between the vortex cores and $R_\mathrm{TF}$ is the TF radius. (c)--(e) $d_\mathrm{sep}$, $R_\mathrm{TF}$, and $d_\mathrm{sep}/R_\mathrm{TF}$ as functions of the particle number $N$ for the stationary vortex dipoles. Crosses in (e) represent the experimental data of Ref.~\cite{Freilich2010}.
}
\end{figure*}

\emph{Results}---Figure~\ref{fig:statdip} summarizes our results for the three-dimensional stationary vortex dipole. For all the particle numbers considered, $3\times 10^5\leq N \leq 7\times 10^5$, the background condensate density has a parabolic TF profile along each coordinate direction. The cores of the vortex and antivortex are observed to be straight, although increasing in size toward the surface of the cloud [Fig.~\ref{fig:statdip}(a)]. In Figs.~\ref{fig:statdip}(c)--\ref{fig:statdip}(e), we plot the separation distance $d_\mathrm{sep}$ between the vortex cores, the TF radius $R_\mathrm{TF}$, and their ratio $d_\mathrm{sep}/R_\mathrm{TF}$ as functions of $N$. Here, $d_\mathrm{sep}$ and $R_\mathrm{TF}$ are determined by fitting the $z$-integrated density $n_\mathrm{int}(x,y)=N\int |\Psi(x,y,z)|^2dz$ to a five-parameter surface consisting of a sum of the parabolic TF profile [Eq.~(\ref{eq:TF})] and two Gaussians; cf. Fig.~\ref{fig:statdip}(b). The separation distance $d_\mathrm{sep}$ is observed to change slowly, decreasing by only $2\,\%$ with $N$ increasing by $133\,\%$. On the other hand, the TF radius $R_\mathrm{TF}$ increases by $16\,\%$, thereby being mostly responsible for the $18\,\%$ reduction in the ratio $d_\mathrm{sep}/R_\mathrm{TF}$. Figure~\ref{fig:statdip}(e) shows that our numerical result for $d_\mathrm{sep}/R_\mathrm{TF}$ is in excellent quantitative agreement with the experimental data of Ref.~\cite{Freilich2010}.

We have also modeled the experiments on the motion of symmetric and asymmetric vortex dipoles. For simplicity, and since no evidence of bent or tilted vortex lines was observed in Ref.~\cite{Freilich2010} [see also Fig.~\ref{fig:statdip}(a)], we have computed the dynamics in two dimensions using Eq.~(\ref{eq:GPE2}). The value of the reduced interaction strength  $g_{\mathrm{2D}}$ is chosen to yield the correct value for the TF radius $R_\mathrm{TF}$. The three-dimensional simulation yields $R_\mathrm{TF}=15.6\ \mu\mathrm{m}$ at $N=6\times 10^5$, from which we obtain $\tilde{g}_\mathrm{2D}=4411$. We have tested the validity of this choice by studying the precession of an off-axis vortex located initially at a radius $r=0.362 R_\mathrm{TF}$ when 5\,$\%$ of the particles are removed every 90 ms starting from $t=0\textrm{ ms}$, and found that the precession frequency increases from $f = 3.15$ Hz at $t=0$ ms to $f =3.61$ Hz at $t=655$ ms, agreeing well with the values $f=3.17$~Hz and $f=3.67$~Hz reported by Freilich \etal \cite{Freilich2010}. The slight discrepancy in $f$ can be partly due to errors in determining the particle number in the experiments.

\begin{figure}
\includegraphics[
  width=210pt,
  keepaspectratio]{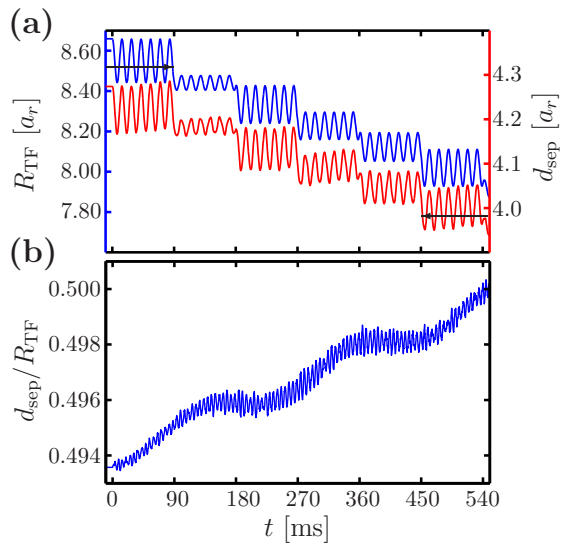}
\caption{\label{fig:statdip-evo} (Color online) Evolution of an initially stationary vortex dipole when $5\,\%$ of the atoms are extracted from the BEC at the time instants indicated on the horizontal axis. Panel (a) shows the Thomas--Fermi radius $R_\mathrm{TF}$ (top blue curve) and the distance $d_\mathrm{sep}$ between the vortices (bottom red curve), whereas panel (b) shows their ratio $d_\mathrm{sep}/R_\mathrm{TF}$.}
\end{figure}

In Ref.~\cite{Freilich2010}, a symmetric vortex dipole was observed to remain stationary for at least 655 ms, during which nine images were extracted from the BEC. To model this experiment, we have computed the evolution of an initially stationary vortex dipole when 5\,\% of the atoms are removed from the trap at 90 ms intervals. The results are presented in Fig.~\ref{fig:statdip-evo}, where $R_\mathrm{TF}$, $d_\mathrm{sep}$, and  $d_\mathrm{sep}/R_\mathrm{TF}$ are plotted as functions of time. In addition to gradually reducing the overall size of the BEC, the atom removal excites the so-called breathing mode (BM), causing  $R_\mathrm{TF}$ and $d_\mathrm{sep}$ to oscillate with the characteristic frequency $f_\mathrm{BM}=2\omega_r /2\pi$ [Fig.~\ref{fig:statdip-evo}(a)]. However, the ratio $d_\mathrm{sep}/R_\mathrm{TF}$ changes by only 1\,$\%$ during the first 540 ms.

\begin{figure}
\begin{center}
\includegraphics[
  width=190pt,
  keepaspectratio]{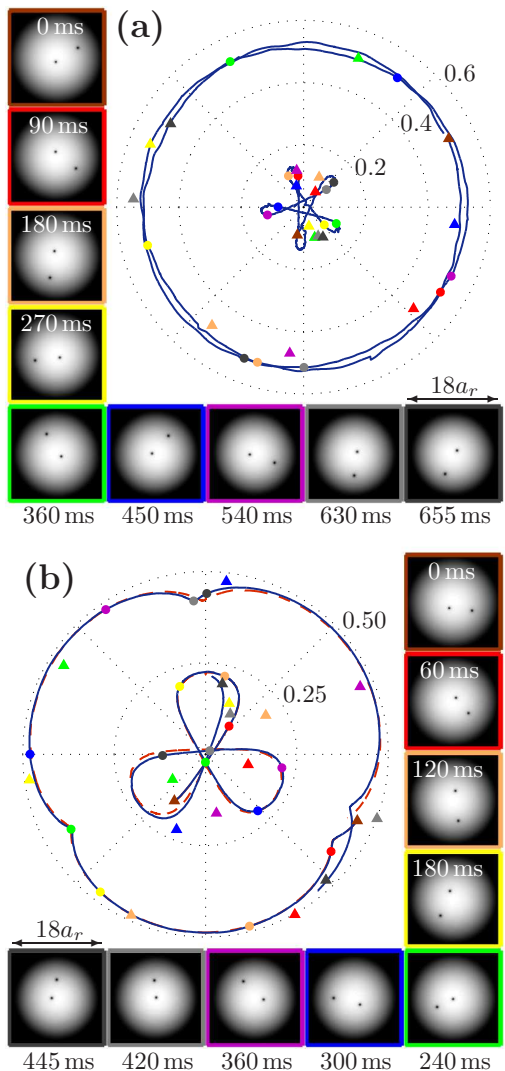}
\end{center}
\caption{\label{fig:traj} (Color online) Trajectories $\left(x_\mathrm{v}/R_\mathrm{TF},y_\mathrm{v}/R_\mathrm{TF}\right)$ of vortices for two different asymmetric vortex dipoles when $5\,\%$ of the atoms are removed every (a) 90 ms or (b) 60 ms from the state (solid blue curves). Dashed red curve in (b) shows the trajectories when no particles are removed. Insets depict the density $|\psi(x,y)|^2$ at each moment of particle extraction; these time instants are indicated by solid circles on the curves. Triangles designate the experimental measurements of Ref.~\cite{Freilich2010} for the same time instants and initial configurations. Marker colors correspond to the image frames.
}
\end{figure}

Freilich \etal \cite{Freilich2010} also studied the precession of asymmetric vortex dipoles, with one vortex near the center of the BEC and the other close to the surface. We have performed a related simulation by first placing the vortex cores at the measured initial positions and then letting the state evolve according to Eq.~(\ref{eq:GPE2}), with $g_{2\mathrm{D}}$ being reduced by the removal of atoms at regular intervals [Eq.~(\ref{eq:g2d})]. The obtained trajectories are presented in Fig.~\ref{fig:traj} together with the corresponding experimental data of Ref.~\cite{Freilich2010}. In Fig.~\ref{fig:traj}(b), we also show the trajectories without the particle removal. We find that the vortices move along approximately periodic trajectories. The atom removal speeds up the vortex motion and excites the breathing mode but does not alter the underlying structure of the paths; it also conserves the orbital angular momentum per atom. The outer (anti)vortex follows a roughly circular trajectory, whereas the inner vortex has a more complicated path with approximate discrete rotational symmetry that is sensitive to the initial vortex positions; see also Ref.~\cite{Middelkamp2010}. The separation distance $d_\mathrm{sep}$ between the vortices oscillates strongly with a period several times shorter than the overall period of precession, which in turn causes the instantaneous speeds of the vortex cores to vary greatly during the evolution. 

\emph{Discussion}---In summary, we have studied the size and dynamics of vortex dipoles in trapped BECs. Contrary to earlier theoretical studies~\cite{Martikainen2001,Schulte2002,Zhou2004,Crasovan2002,Crasovan2003,Mottonen2005,Pietila2006,Middelkamp2010,Stockhofe2010,Li2008,Klein2007,Geurts2008}, which have considered only pancake-shaped condensates or small particle numbers, we found stationary vortex dipoles in a three-dimensional configuration in the TF limit corresponding to large atomic clouds. In addition, we investigated the evolution of nonstationary vortex dipoles, and found that the vortices follow complicated trajectories whose exact shapes depend strongly on the initial vortex positions. 

This study explains the recent experiment by Freilich \etal \cite{Freilich2010} on the real-time dynamics of vortex dipoles. Our numerical results for the size of stationary vortex dipoles are in excellent agreement with their experimental data. We also studied how the atom 
loss due to the experimental imaging method affects the vortex dynamics: The major effect is the excitation of the breathing mode of the whole BEC, which can be subtracted from the data at will by a global distance scaling. Otherwise, the vortex trajectories are altered only slightly. Although in some cases there may exist more subtle issues such as the onset of dynamical instabilities, the multishot imaging method used by Freilich \etal appears to be a promising tool for studying vortex dynamics in trapped BECs.

\begin{acknowledgments}
The authors thank the Academy of Finland, the Emil Aaltonen Foundation, and the V\"ais\"al\"a Foundation for financial support. David S. Hall is acknowledged for providing the measurement data and details of the experiment.
\end{acknowledgments}

\bibliography{vortexdipole}

\end{document}